\documentclass{emulateapj}

\shorttitle{Pulsar Force-free Magnetosphere}
\shortauthors{Petrova}

\begin{document}

\title{On the Global Structure of Pulsar Force-free Magnetosphere}

\author{S. A. Petrova}
\affil{Institute of Radio Astronomy, Chervonopraporna
Str., 4, Kharkov 61002, Ukraine}
\email{petrova@ri.kharkov.ua}

\begin{abstract}
The dipolar magnetic field structure of the neutron star is modified by the plasma originating in the pulsar magnetopshere. In the simplest case of a stationary axisymmetric force-free magnetosphere, a self-consistent description of the fields and currents is given by the well-known pulsar equation. Here we revise the commonly used boundary conditions of the problem in order to incorporate the plasma-producing gaps and to provide a framework for a truly self-consistent treatment of the pulsar magnetosphere. The generalized multipolar solution of the pulsar equation is found, which, as compared to the customary split monopole solution, is suggested to better represent the character of the dipolar force-free field at large distances. In particular, the outer gap location entirely inside the light cylinder implies that beyond the light cylinder the null and critical lines should be aligned and go parallel to the equator at a certain altitude. Our scheme of the pulsar force-free magnetosphere, which will hopefully be followed by extensive analytic and numerical studies, may have numerous implications for different fields of pulsar research. 
\end{abstract}
\keywords{pulsars: general --- stars: magnetic field --- stars: neutron --- plasmas --- magnetohydrodynamics}

\section{Introduction}

Pulsars are the rotating magnetized neutron stars \citep{h68} with the rotation periods of $~10^{-3}-1$~s and the surface magnetic field strengths of $~10^9-10^{12}$~G. The induction electric field can extract charged particles from the stellar surface and supply the pulsar magnetosphere with plasma \citep{gj69}. In the tube of open magnetic field lines, the longitudinal electric field accelerates the particles to the energies of $~10^{12}-10^{13}$~eV, enabling them to launch pair cascades \citep{rs75}. The resultant secondary electron-positron plasma screens the accelerating field all over the tube, except for several small regions of pair production called gaps (see Fig.~\ref{f1}a, \citealt{rs75,a83,chr86}). The closed field lines are equipotential, and the primary plasma they bear screens the accelerating longitudinal electric field everywhere.

The presence of the plasma modifies the original dipolar structure of the pulsar magnetic field. The problem of a self-consistent description of fields and currents in the pulsar magnetosphere was formulated in the form of a well-known pulsar equation \citep{m73,sw73,o74}. The basic model is that of a stationary axisymmetric force-free dipole, where the magnetic and rotational axes are aligned, the electromagnetic forces are balanced and the particle inretia is ignored. An exact analytical solution of the pulsar equation was found only for the magnetic monopole located at the origin \citep{m73}, in which case the field remains monopolar everywhere. In the case of a magnetic dipole, the pulsar equation was solved in terms of series for several specific forms of the current distribution \citep{m73b,bgi83}. However, these solutions are valid only inside the light cylinder and cannot be smoothly continued to the infinity.

A certain progress was achieved in the pioneering work of \citet{ckf99}, which was devoted to the numerical treatment of the pulsar equation. As a part of the numerical procedure, the current distribution was constructed so as to provide a smooth behaviour of the magnetic field lines all over the space. The numerical solution of \citet{ckf99} was subsequently reproduced by means of other algorithms \citep{g05,k06,mck06,s06} and generalized in a number of aspects. In particular, the non-axisymmetric case was addressed \citep{s06,kc09,bs10,kck12}, the polar gap potential drop and the resultant differential rotation of the magnetosphere were included \citep{c05,t06,t07}, the non-ideal magnetosphere with a finite conductivity was considered \citep{kkhc12,lst12}. All these studies confirmed the formal validity of the original results obtained by \citet{ckf99}. However, it was soon recognized \citep{t06} that the polar gap can hardly be responsible for the numerically simulated current distribution, in which the reverse current partially flows on the open field lines controlled by the gap. This questions the very existence of the stationary force-free configuration in pulsars and challenges both the polar gap theories and the pulsar magnetosphere models. The difficulty of including the polar gap into the global structure of the pulsar magnetosphere stimulated numerous studies of the non-stationary polar gap \citep{lev05,melr08,l09,t10}.
In the present paper, we develop another approach to solving this difficulty. Namely, we revise the physical model underlying the numerical simulations of the pulsar force-free magnetosphere.

We turn to the analytical treatment of the pulsar equation and find the multipolar solution, which generalises the monopolar one. Based on this result, we suggest a new model of the dipolar force-free magnetosphere of a pulsar, which includes all the magnetospheric gaps into both the boundary conditions and the current circuit configuration. Section~2 contains the basics of the existing model. In Sect.~3, we search for the multipolar solutions of the pulsar equation. An improved geometrical model of a force-free dipole beyond the light cylinder is constructed in Sect.~4. The force-free magnetosphere, which properly includes the magnetosphereic gaps, is described in Sect.~5. Our results are briefly summarized and discussed in Sect.~6.

\section{Basic equations} 
\protect\label{s2}

A starting point for the studies of pulsar electrodynamics is the model of a stationary axisymmetric force-free magnetosphere. It is convenient to choose the cylindrical coordinate system $(\rho,\phi,z)$ with the axis along the pulsar axis. Then the pulsar equation reads \citep{m73,sw73,o74} 
\begin{equation}
\left (1-\rho^2\right )\left[\frac{\partial^2f}{\partial\rho^2}+\frac{1}{\rho}\frac{\partial f}{\partial\rho}+\frac{\partial^2f}{\partial z^2}\right]-\frac{2}{\rho}\frac{\partial f}{\partial\rho}=-A\frac{\mathrm{d}A}{\mathrm{d}f},
\label{eq4}
\end{equation}
where the dimensionless functions $f(\rho,z)$ and $A(f)$ are proportional respectively to the magnetic flux and electric current through the circle of a radius $\rho$ centered at the magnetic axis at an altitude $z$ above the origin. Both functions are unknown, and the problem lies in finding the current distribution which makes the self-consistent magnetic field obey the physically meaningful boundary conditions.

\begin{figure}
\vspace{20mm}
\includegraphics[width=50mm]{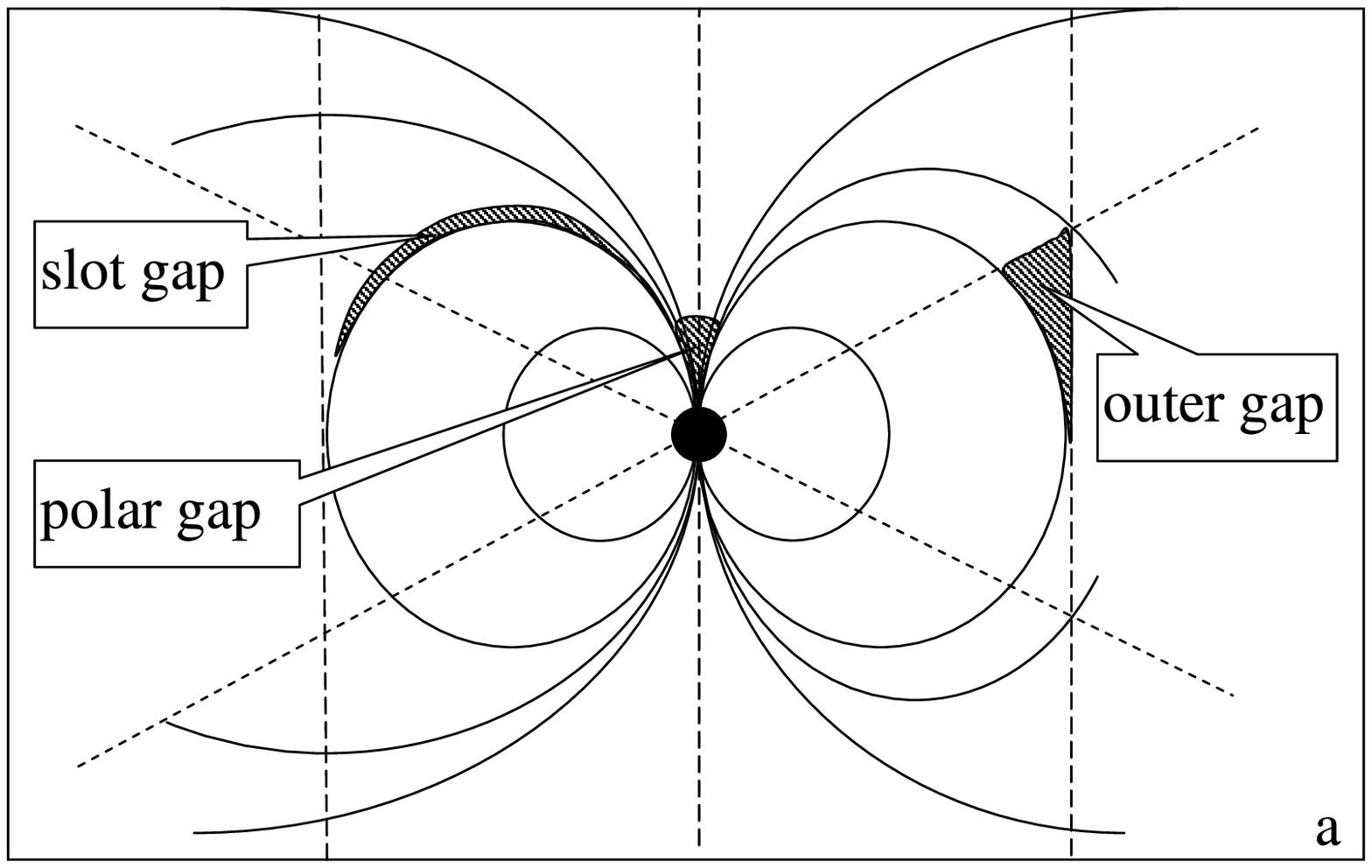}

\vspace{10mm}

\includegraphics[width=50mm]{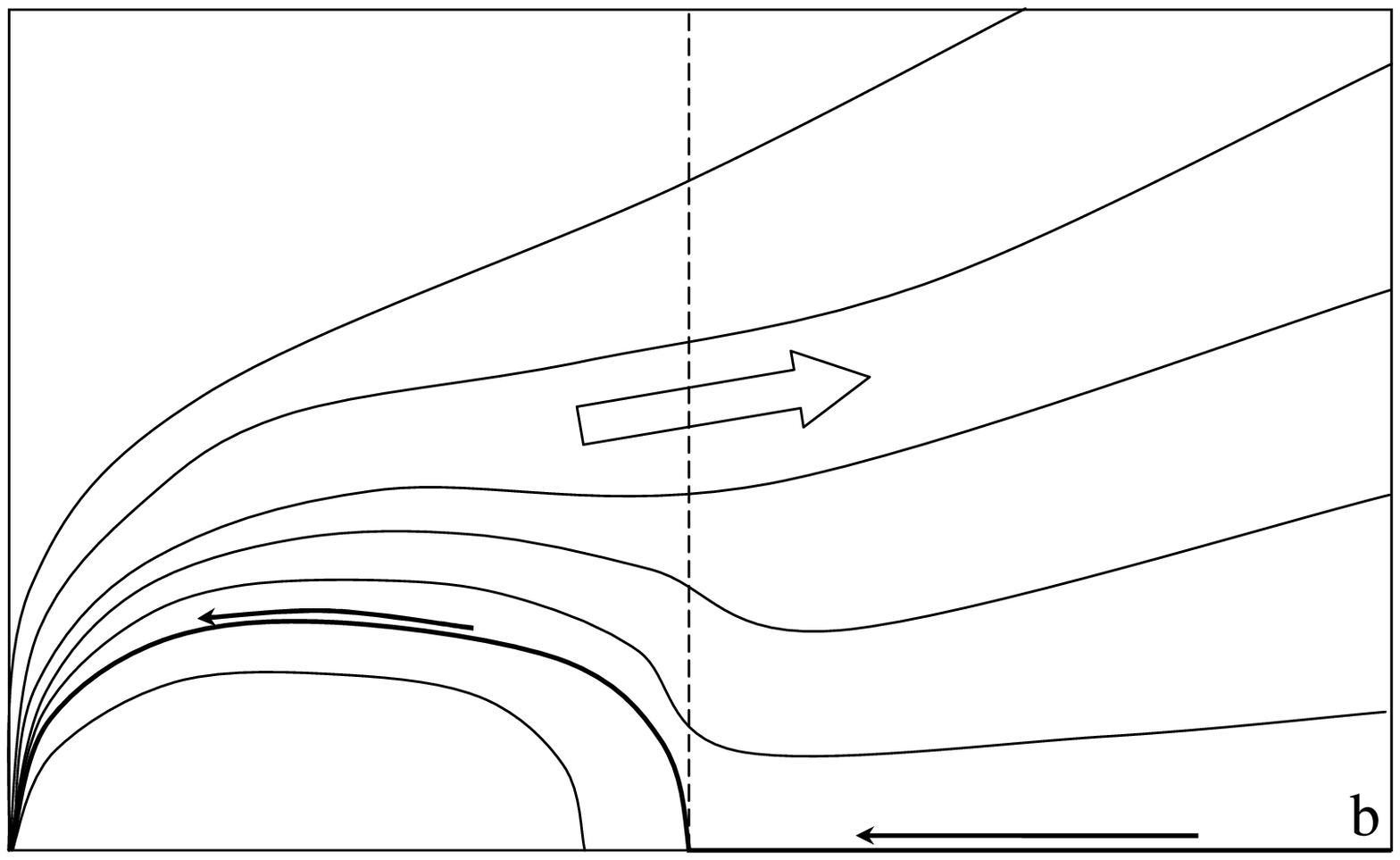}

\vspace{10mm}
\includegraphics[width=50mm]{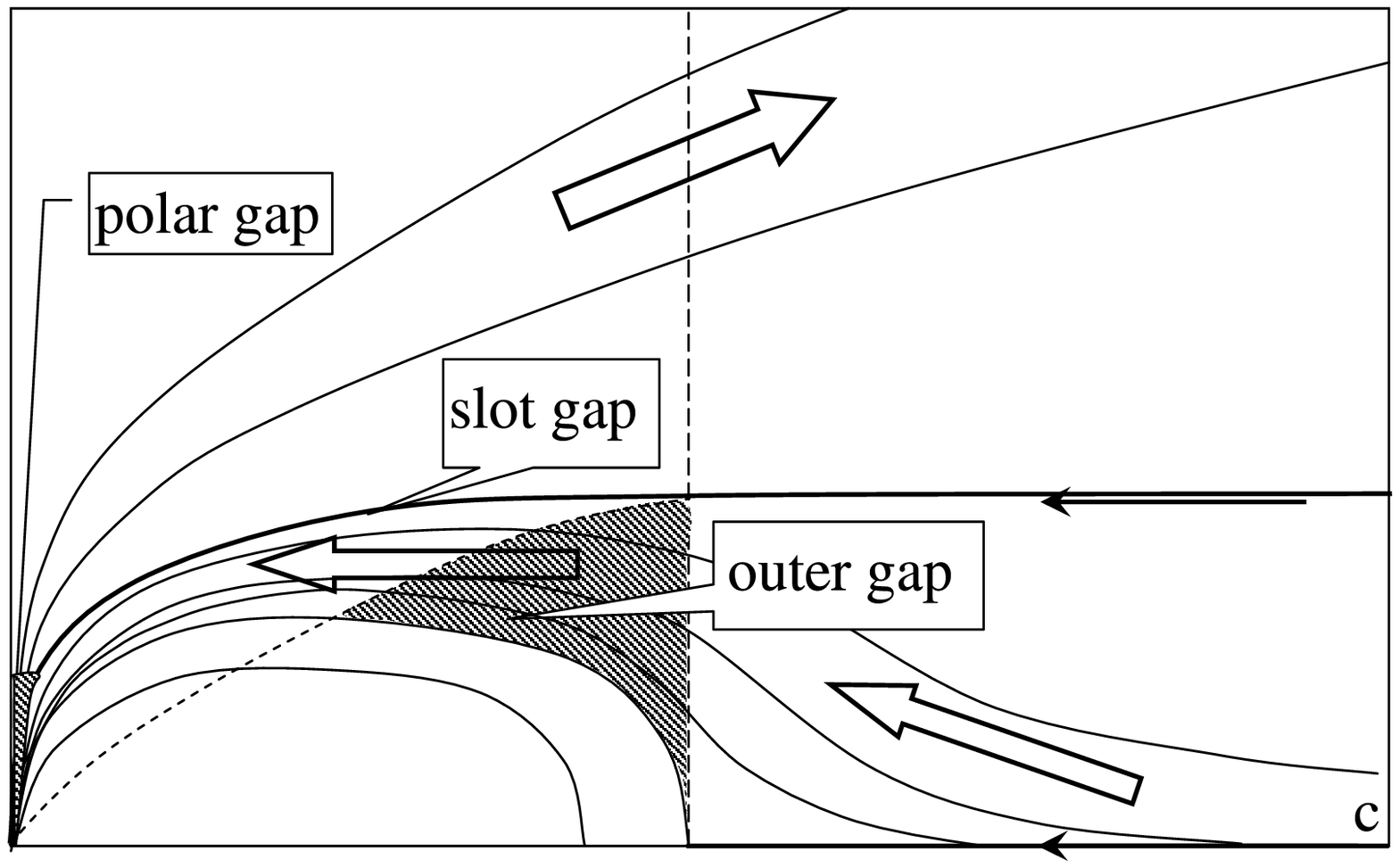}
\caption{Schematic structure of pulsar magnetosphere: {\bf a} -- the case of an axisymmetric vacuum rotating dipole; {\bf b} -- the existing force-free model; {\bf c} -- the model suggested in the paper. Solid lines represent the magnetic field lines; dashed lines show the light cylinder boundary, where the particle rotational velocity equals the speed of light; dotted lines delineate the null line, where the charge density necessary to screen the accelerating electric field changes the sign; thick lines correspond the the critical line, where the current density becomes zero; the direction of current is shown by arrows.}
\label{f1}
\end{figure}

Strictly speaking, an exact form of the boundary conditions is not known in advance as well. The set of assumptions usually taken in the numerical simulations is schematically presented in Fig.~\ref{f1}b. At the neutron star surface, the magnetic field is dipolar, $f=\rho^2/(\rho^2+z^2)^{3/2}$; at infinity, the field lines become radial, $f=f(\rho/z)$; beyond the light cylinder, the separatrix between the open and closed field line regions goes along the equator, $\partial f/\partial\rho=0$; the closed field lines cross the equator perpendicularly, $\partial f/\partial z=0$. The current function $A\mathrm{d}A/\mathrm{d}f$ is zero on the closed field lines and at the magnetic axis; the return current flows along the separatrix and on the neighbouring open field lines. It should be noted that this standard set of assumptions does not allow for the presence of the plasma-producing gaps in the pulsar magnetopshere, and the magnetospheric structure derived on its basis cannot be completely self-consistent. The question of a proper inclusion of the gaps into the global magnetospheric structure will be addressed below.

\begin{figure}
\includegraphics[width=70mm]{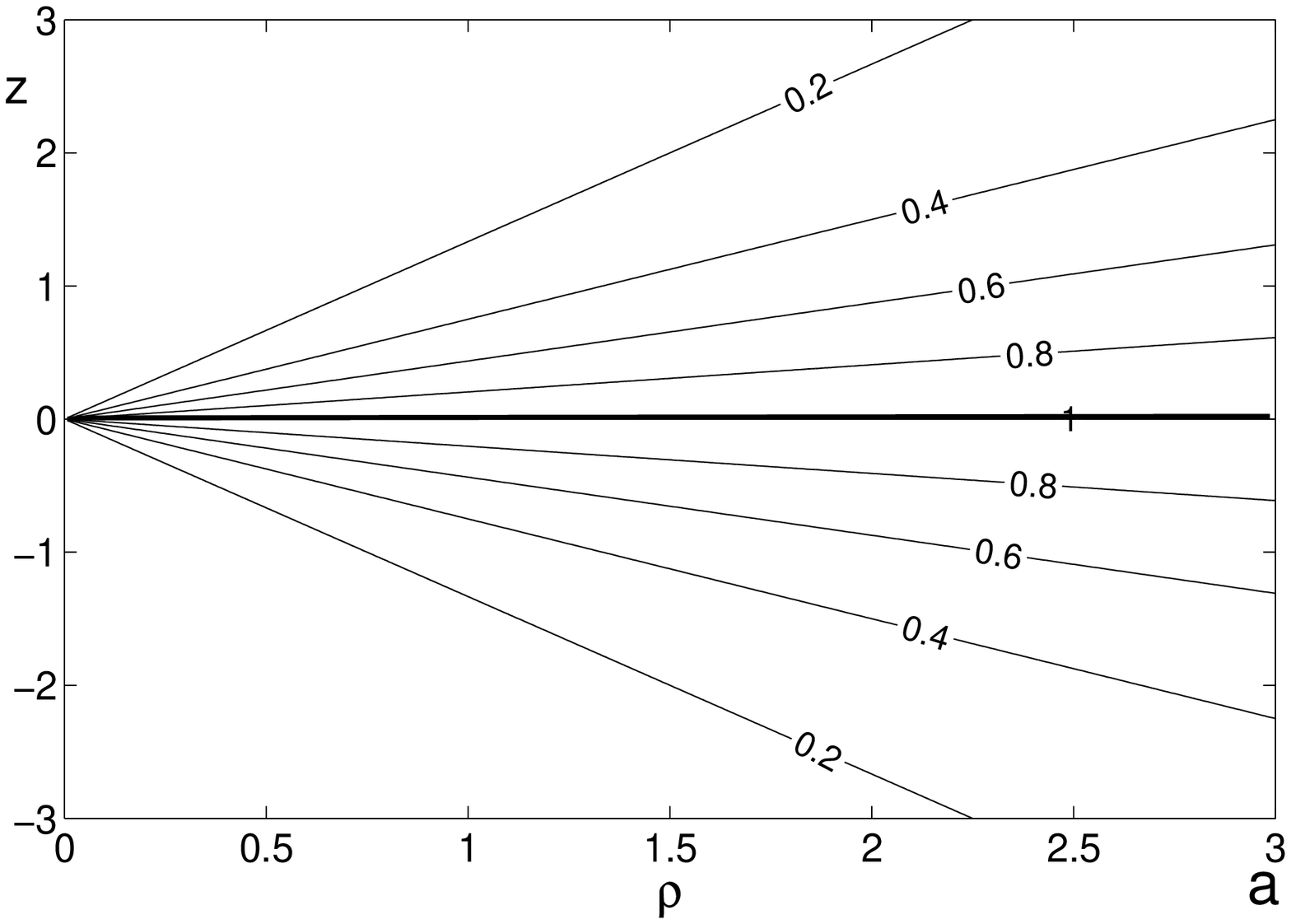}
\includegraphics[width=70mm]{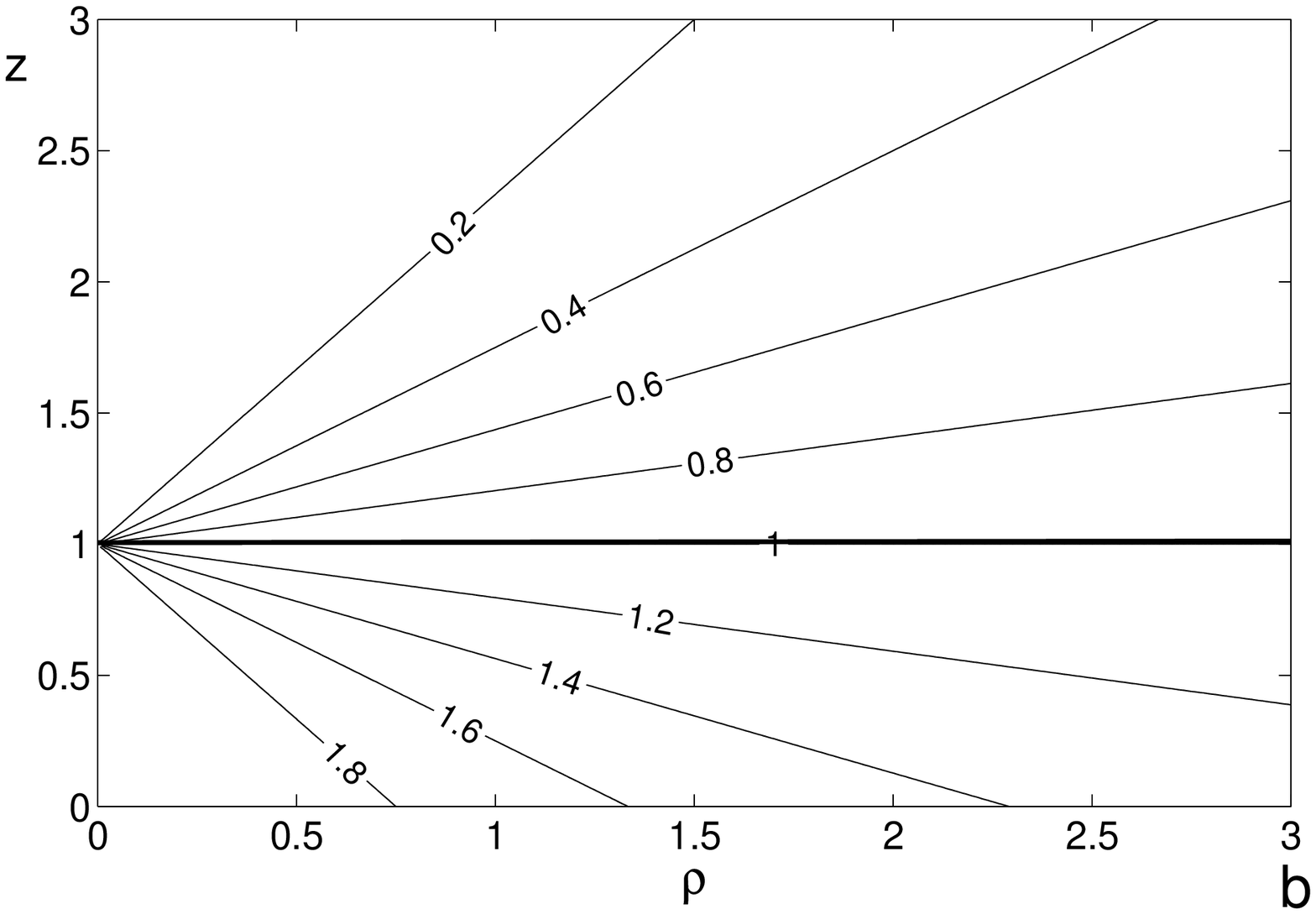}
\includegraphics[width=70mm]{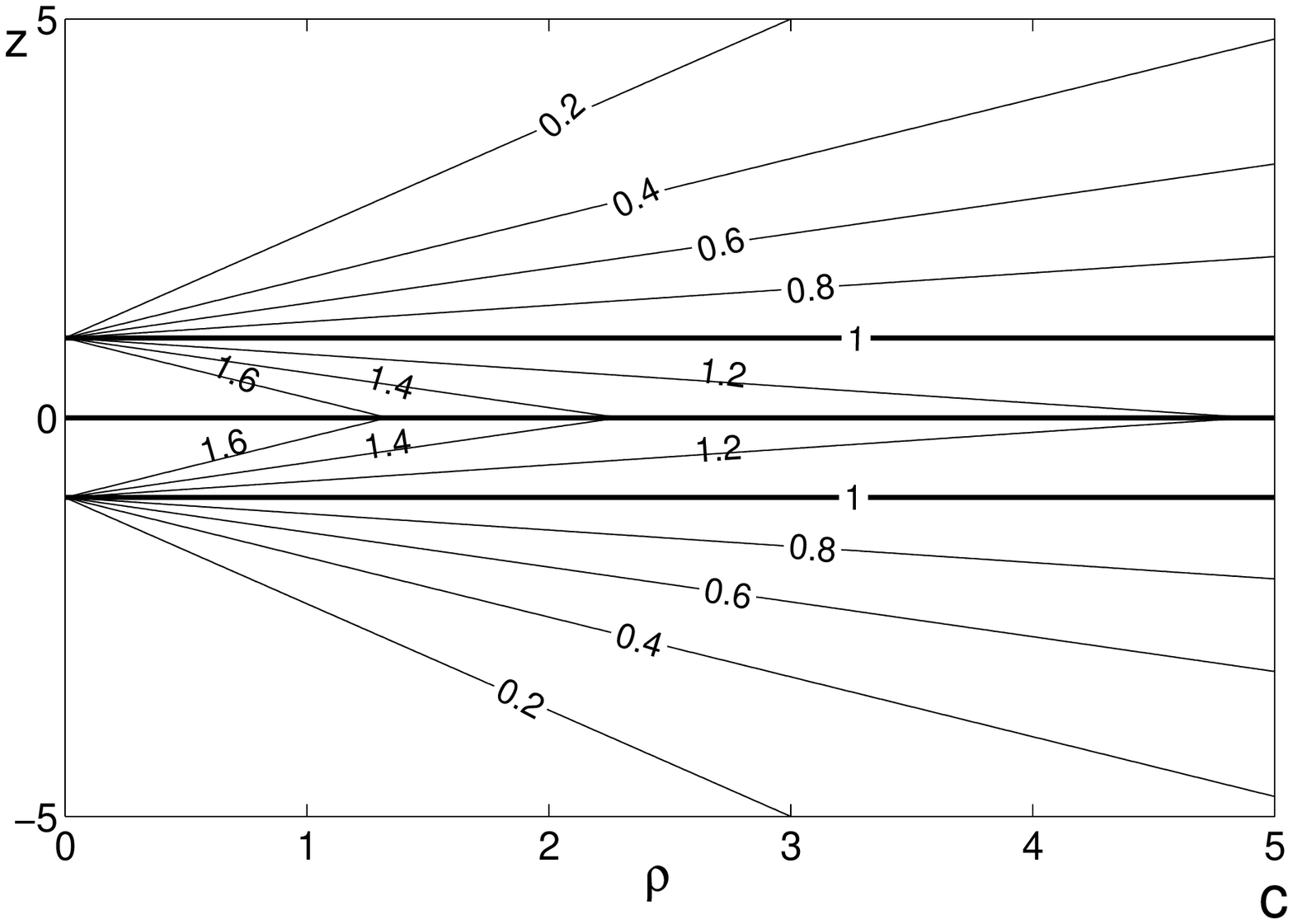}
\caption{Subsidiary force-free configurations: {\bf a} -- the model of a centered split monopole. The level lines of the flux function $f$ are shown together with the level values. {\bf b} -- the field of an offset monopole according to Eq.~(\ref{eq7}); $f_0=1$, $a=1$. {\bf c} -- the split-offset monopole model; $f_0=1$, $a=1$. The critical field lines and the coincident null lines are shown by thick lines.}
\label{f2}
\end{figure}

\section{Exact solutions of the pulsar equation}
\protect\label{s3}

An exact analytical solution of the pulsar equation (\ref{eq4}) is known only for the case of a magnetic monopole located at the center of a neutron star. In order to generalize this result, we searched for the multipolar solutions, which satisfy the relation
\begin{equation}
\frac{\partial^2f}{\partial\rho^2}-\frac{1}{\rho}\frac{\partial f}{\partial\rho}+\frac{\partial^2f}{\partial z^2}=0.
\label{eq5}
\end{equation}
Then Eq.~(\ref{eq4}) turns into
\begin{equation}
2\rho\frac{\partial f}{\partial\rho}=A\frac{\mathrm{d}A}{\mathrm{d}f}.
\label{eq6}
\end{equation}
Based on Eq.~(\ref{eq6}), the magnetic flux function can be presented as $f(\rho,z)=f[\rho\xi(z)]$, where $\xi(z)$ is an arbitrary function. Using this in Eq.~(\ref{eq5}) yields
\begin{displaymath}
f=f_0\rho^2(z-a),\quad A=2f
\end{displaymath}
and
\begin{equation}
f=f_0\left[1-\frac{z-a}{\sqrt{(z-a)^2+\rho^2}}\right],\quad A=f\left(2-\frac{f}{f_0}\right),
\label{eq7}
\end{equation}
where $f_0$ and $a$ are arbitrary constants. The solution given by Eq.~(\ref{eq7}) is of particular interest. It represents the field of a monopole offset along the $z$-axis by a distance $a$ (see Fig.~\ref{f2}b), with the current function being identical to that of a monopole centered at the origin. From the mathematical point of view, such a generalization of the monopole solution could be directly expected from the form of Eq.~(\ref{eq4}), since it does not depend on the $z$-coordinate explicitly and a shift along this coordinate changes nothing. The implications of the offset monopole solution for the magnetosphere of a force-free dipole are discussed in Sect.~\ref{s4}.

In the spherical coordinate system $(r,\theta,\varphi)$, where $r=\sqrt{\rho^2+z^2}$ and $\theta=\mathrm{atan}(\rho/z)$, the solution (\ref{eq7}) can be presented as an infinite sum of the centered magnetic multipoles,
\begin{eqnarray}
\frac{f}{f_0}=1+\frac{1-(r/a)\cos\theta}{\sqrt{1-2(r/a)\cos\theta+r^2/a^2}}=1-\cos\theta\nonumber \\
+\sum_{k=1}^\infty\left (\frac{a}{r} \right )^k\left[P_{k-1}(\cos\theta)-\cos\theta P_k(\cos\theta)\right],\, \frac{r}{a}>1,\nonumber
\end{eqnarray}
where $P_k(\cos\theta)$ are the Legendre polynomials. This result can be directly obtained from the pulsar equation written in the spherical coordinate system by expanding the solution into a series in $1/r^k$ and taking into account Eq.~(\ref{eq5}). Note also that although the pulsar equation is non-linear, the polynomial function in $f$, standing on the right-hand side, is also a series in $1/r^k$. At $r\to\infty$, the solution coincides with that for the centered monopole. But at finite distances the critical line, $f=f_0$, goes parallel to the equator at an altitude $a$ (cf. Fig.~\ref{f2}b).

\section{Geometrical model of a force-free dipole beyond the light cylinder}
\protect\label{s4}

Although the assumption of the neutron star field in the form of a monopole is non-realistic, it is generally believed that the version of a split monopole (see Fig.~\ref{f2}a), where the solution in the upper half-plane is symmetrically continued beyond the magnetic equator, well mimics the structure of the force-free dipole at infinity \citep{m91}. However, at finite distances it does not allow for the consequences of the essentially dipolar features, such as the presence of the magnetospheric gaps and the closed field line region. This will be amended in our geometrical scheme based on the offset monopole.

The analogy with an offset monopole enables to account for the presence of the outer gap in the pulsar magnetosphere. This gap forms at the intersection of the open field lines with the null line \citep{chr86}, along which the magnetospheric charge density \citep[e.g.][]{ckf99}
\begin{equation}
\rho_e=\frac{\Omega}{4\pi c}\frac{A\mathrm{d}A/\mathrm{d}f-(2/\rho)\partial f/\partial\rho}{1-\rho^2}
\label{eq8}
\end{equation}
equals zero. It is natural to assume that the outer gap is located entirely within the light cylinder. In order to provide a proper outer gap location, the null line should go along a certain field line all the way beyond the light cylinder. Furthermore, if one take that the outer gap produces the reverse current necessary to close the pulsar circuit, then the null line should coincide with the critical line, at which the current function $A\mathrm{d}A/\mathrm{d}f$ equals zero (see Fig.~\ref{f1}c). Finally, these two lines should be parallel to the equator (cf. Eq.~(\ref{eq8})), similarly to the case of an offset monopole (see Fig.~\ref{f2}b). Therefore we suggest that the force-free field of a dipole at large distances is better represented by the split-offset monopole scheme based on the two symmetrically offset monopoles of opposite polarity (see Fig.~\ref{f2}c).

Further generalization of the monopolar case can be made from analysing the pulsar equation which includes differential rotation \citep[see, e.g.,][]{c05},
\begin{eqnarray}
\left (1-\rho^2\Omega^2\right )\left[\frac{\partial^2 f}{\partial \rho^2}+\frac{1}{\rho}\frac{\partial f}{\partial\rho}+\frac{\partial^2 f}{\partial z^2}\right]-\frac{2}{\rho}\frac{\partial f}{\partial\rho} \nonumber \\
=-A\frac{\mathrm{d}A}{\mathrm{d}f}+\rho^2\Omega\frac{\mathrm{d}\Omega}{\mathrm{d}f}\left[\left(\frac{\partial f}{\partial\rho}\right)^2+\left(\frac{\partial f}{\partial z}\right)^2\right],
\label{eq9}
\end{eqnarray}
where $\Omega=\Omega(f)$ is the angular velocity of the magnetosphere rotation allowing for the potential drop across the open magnetic field lines. Integration of Eq.~(\ref{eq9}) {\it for the case of a monopole} yields
\begin{equation}
A=\Omega f\left(2-f/f_0\right).
\label{eq10}
\end{equation} 
It is interesting to examine the split-offset monopole configuration for different pairs $(A,\Omega)$ obeying Eq.~(\ref{eq10}). As the magnetic and electric field strengths read
\begin{equation}
{\bf B}=\frac{1}{\rho}\left(-\frac{\partial f}{\partial z},A,\frac{\partial f}{\partial\rho}\right),\, {\bf E}=-\Omega\left(\frac{\partial f}{\partial \rho},0,\frac{\partial f}{\partial z}\right),
\label{eq11}
\end{equation}
one can see that along the equator
\begin{equation}
B^2-E^2=\frac{f_0^2}{(a^2+\rho^2)^2}
\label{eq12}
\end{equation}
and along the horizontal null (critical) lines, $f=f_0$,
\begin{equation}
B^2-E^2=f_0^2/\rho^4,
\label{eq13}
\end{equation}
i.e. in all these cases $B^2-E^2=B_p^2>0$ (where $B_p$ is the poloidal field) and the force-free approximation is valid for any pair $(A,\Omega)$. Moreover, as the right-hand sides of Eqs.~(\ref{eq12})-(\ref{eq13}) are independent of $z$, the equilibrium condition for the three above mentioned horizontal lines, $\mathrm{d}(B^2-E^2)/\mathrm{d}z=0$ \citep{o74,l90}, is also fulfilled for any $(A,\Omega)$. Thus, the four regions bounded by the three horizontal lines in Fig.~\ref{f2}c may have different $(A,\Omega)$.

The split-monopole scheme contains from one to three current sheets located along the null (critical) lines and equator. In the simplest case, $A=\Omega=0$ in the equatorial region between the lines $f=f_0$ and $A\neq 0$ between the magnetic axis and these lines. Then the symmetric current sheets along the null (critical) lines close the current circuit in each hemisphere. Given that in the equatorial region $A\neq 0$ as well, the regions on both sides of the null (critical) lines may join without a current sheet, in which case the return current flows along the equator and the equatorial field lines entering the equatorial current sheet. Note that a similar structure of the equatorial region beyond the light cylinder is characteristic of the dipolar force-free magnetosphere simulated in \citet{g11a,g11b}. The only distinction is that in the dipolar case $B^2-E^2=0$ along the equator. It should be kept in mind that the realistic magnetosphere of a pulsar may bear not only quantitative but also qualitative distinctions from the simplistic monopole-based scheme. In particular, a simpler magnetic field structure in the equatorial region, which may result from reconnections, is not excluded. Nevertheless, the split-offset monopole scheme suggested above is believed to give insight into the general structure of the pulsar magnetosphere allowing for the coexisting gaps.

\section{New model of dipolar force-free magnetopshere}
\protect\label{s5}

Based on the split-offset monopole model, the global structure of the stationary axisymmetric dipolar force-free magnetosphere of a pulsar seems to look as follows (see Fig.~\ref{f1}c). The critical field line, which becomes parallel to the equator beyond the light cylinder, divides the overall open field line region into the two parts. The upper part, where the field lines are inclined to the magnetic axis at an acute angle all the way up to infinity, is controlled by the polar gap. The lower part, where the field lines ultimately enter the equator, is controlled by the outer gap. The two gaps are adjusted by the slot gap located between them. Note that a similar configuration of the gaps acting at different bundles of open magnetic field lines was recently obtained in \citet{s12} by means of numerical simulations of the plasma particle motions in the pulsar magnetosphere.

The current circuit looks as follows. The direct current flows along the field lines controlled by the polar and slot gaps and returns to the star in the equatorial current sheet, which stretches up to the light cylinder, along the equatorial filed lines and through the outer gap. Alternatively, beyond the light cylinder the current may return in the current sheet along the null (critical) line and inside the light cylinder through the outer gap. In this case, the longitudinal current flowing through the outer and slot gaps changes along the magnetic lines. Such a change may be compensated for by the trans-field currents in the vicinity of the light cylinder, so that continuity equation may still be satisfied.

Presumably, the outer gap is not infinitesimally narrow, in which case the force-free consideration is no longer appropriate. Moreover, the equatorial region beyond the light cylinder is suggestive of the dissipation processes, which should be studied out of the framework of the force-free approximation in conjunction with the outer gap physics. It is important to note that there may be a physical relation between the upper and lower hemispheres through the outer-gap-controlled field lines. We also anticipate that the non-stationary dissipation processes in the equatorial region may be responsible for the magnetar-like activity of pulsars.

\section{Discussion and conclusions}
\protect\label{s6}

We have found a new exact analytic solution of the pulsar equation, which generalizes the well-known monopolar solution to the case of an offset monopole and presents an infinite series over the centered multipoles. We argue that, as compared to the classical split monopole, the geometrical model based on the two symmetrically offset monopoles of opposite polarity better mimics the character of the dipolar force-free field beyond the light cylinder. In particular, it allows for the presence of the outer gap in the pulsar magnetosphere. In our model, the polar and outer gaps control different bundles of open magnetic field lines, the two gaps being adjusted by the slot gap located between them. Then the direct current flows through the polar and slot gaps and returns to the neutron star through the outer gap. Generally speaking, in the outer and slot gaps the longitudinal current may change along a field line because of the trans-field currents in the vicinity of the light cylinder.

The main distinction of our force-free model from the customary one \citep[e.g.,][]{ckf99} lies in the equatorial magnetic field structure. In this respect, our model is similar to that of \citet{g11a,g11b}, who was the first to suggest the alternative boundary condition at the equator based on his simulations in the context of strong-field electrodynamics. Of course, those works did not incorporate the plasma-producing gaps into the magnetospheric structure, but it is the attempt to allow for the finite conductivity that have demonstrated the necessity to reconsider the equatorial boundary condition.

Our schematic model of the pulsar force-free magnetosphere including the magnetospheric gaps is believed to be a proper framework for detailed analytic and numerical studies. In particular, we plan to give an exact analytic description in the forthcoming papers. The first paper of the series \citep{p12} concentrates on the region close to the magnetic axis, and the main result is that the force-free field at the top of the polar gap differs from that of a vacuum dipole. This is attributed to the action of the transverse current which flows at the neutron star surface and closes the pulsar current circuit.

Our model may also give insight into a detailed picture of the particle flows supporting the necessary global currents and into the underlying physics of the magnetospheric gaps. Being brought into correspondence with the global magnetospheric structure, geometry and physics of the gaps should have weighty implications for the properties of both the high-energy emission and the secondary plasma produced there. It will become possible to finally establish which gap is responsible for the observed non-thermal high-energy emission and what is the underlying physical mechanism. This is especially important in the light of the recent progress in pulsar studies at GeV energies \citep[e.g.,][]{abdo10}. The refined secondary plasma characteristics would be useful for elaborating the theory of radio wave propagation in the magnetosphere \citep{pl00} and may provide a clue for understanding the pulsar radio emission mechanism.

The model of a stationary force-free magnetosphere can be regarded as a starting point for studying the instabilities in the plasma flow and the resultant diversiform fluctuations in pulsar radio emission over a wide range of timescales, including those yet to be discovered in the present and future pulsar transient searches at the world largest radio telescopes \citep{gurt,lofar,ska}. Analysis of the pulsar energy losses related not only to the current losses on the open field lines, but also to the energy release via reconnections \citep{l96} and outbursts \citep{c05} in the equatorial region, would uncover the relevant scenarios of the neutron star spin evolution, cast new light on the magnetar-like activity \citep{mereghetti} and enable one to construct a physically grounded classification of the neutron star observational manifestations beyond the framework of the classical rotating radio pulsar concept \citep{kaspi}.

\acknowledgements
I am grateful to Prof. S. Shibata for useful discussion.
The work is partially supported by the grant of the President of Ukraine (the project of the State Fund for Fundamental Research No. F35/554-2011).

\end{document}